\documentclass[%
 aip,
 jmp,%
 amsmath,amssymb,
 reprint,%
]{revtex4-2}

\usepackage{graphicx}
\usepackage{dcolumn}
\usepackage{bm}
\begin{document}

\preprint{AIP/123-QED}

\title{Resource competition in Three-gene-motif \& Emergence of Feed-forward response: A Spatiotemporal Study}
\author{Priya Chakraborty}
\email{pc.20ph1104@phd.nitdgp.ac.in}
\affiliation{Department of Physics, National Institute of Technology Durgapur, West Bengal,  India}
\author{Ushasi Roy}
\email{ushasi@iiserpune.ac.in}
\affiliation
{Department of Physics, Indian Institute of Science Education and Research, Pune, India}
\author{Sayantari Ghosh}
\email{sayantari.ghosh@phy.nitdgp.ac.in}
\affiliation{Department of Physics, National Institute of Technology Durgapur, West Bengal,  India}


\date{\today}

\begin{abstract}
Feed-forward dynamics, which is well-known to have several important implications in nonlinear dynamical systems, frequently occurs in gene expression motifs, and has been well explored experimentally and mathematically. 
However, dependency of the components of a genetic circuit upon its host, due to the requirement for resources like ribosome, ATP, transcription factors, tRNA, etc., and related effects are of utmost importance, which is commonly ignored in mathematical models. In a resource-limited environment, two apparently unconnected genes can compete for resources for their respective expression and may exhibit indirect regulatory connection; an emergent response thus arises in the system completely because of resource competition.  In this work, we have shown how the responses of the feed-forward loop (FFL), a well-studied regulatory genetic motif, can be recreated considering the resource competition in a three-gene pathway. Exploring the genetic system with temporal as well as spatiotemporal stability analysis, interesting transient and steady-state responses have been observed. The genetic motifs explored in this paper show many of the characteristic features of the conventional FFL structure, like response delay and pulse generation. Most interestingly, in a two-dimensional cellular arrangement,  characteristic pattern formation under a concentration gradient of input signals have also been observed. This study pinpoints a larger area of research and exploration in synthetic and cellular systems, which will reveal novel controlling ideas and unique behavioral changes in the system for its context dependencies.
\end{abstract}

\keywords{Resource competition, Feed-forward loop, Pattern formation}
\maketitle
\section{Introduction}
\label{intro}
Naturally occurring dynamical systems are nonlinear, and specific architectures or interactions drive these systems into interesting time evolution and steady state responses. One way to greatly enhance the performance of a dynamic system is through feed-forward control \cite{hu2014feedforward}. A system's response times, oscillations, and overall performance can all be improved by adding feed-forward structure. The system can become more responsive and stable by using feed-forward control, which can detect and correct disturbances before they have an impact on it. Feed-forward loop (FFL) is one of the most abundantly found structures in gene regulatory networks inside living cells. Living cells carry some re-occurring genetic subgraphs called \textit{motifs}, which are responsible for some key functions of cell physiology \cite{milo2002network,shen2002network}. Existence of FFL motif in yeast \cite{shen2002network}, \textit{C. elegence}  \cite{mangan2003coherent}, \textit{B. subtilis} \cite{milo2004superfamilies,eichenberger2004program}, Sea urchin \cite{milo2004superfamilies},  \textit{E. coli} \cite{milo2004superfamilies,mangan2003coherent}, fruit fly \cite{milo2004superfamilies}, human \cite{odom2004control} and in many more diverse organisms are already seen. Exhaustive studies have been performed to explore the different possible perspectives of this motif in living systems, like, occurrence and fundamental responses, global relative parameter sensitivities, cost-benefit analysis, relationships between noise, functionality, stochastic resonance etc.\cite{alon2007network,ghosh2005noise,dekel2005environmental,guo2009stochastic,wang2012global}. \\ Focusing on many such motifs, including FFL, synthetic biologists have designed models, first mathematically and then experimentally, to understand the nature of the dynamics in whole-cell or organism scenarios. The synthetically constructed gene arrangements, often referred to as the \textit{circuits}, are implemented in living cells, called the \textit{host} for operating with particular feature tasks. Interestingly the dynamics of the host and circuits often get coupled, most of the time in a nonlinear fashion. The functionality of the implemented synthetic circuit depends upon its host's physiology, host growth and replication, operating temperature, pH, binding specificities, and several other factors. Recent studies provide sufficient information that the host's physiology can affect, modulate or modify the circuit response immensely. On a low-scale modulation, in terms of parameter regime, or in a high-scale modulation, the output response can be completely unpredictable, giving rise to a new response or even a complete failure of the experiment \cite{weisse2015mechanistic,tan2009emergent,towbin2017optimality}. Limited knowledge about inter-cellular dynamics may restrict us from explaining these emergent responses, arising mainly because of the nonlinear couplings between the host and the circuit. Over the last decade, researchers have started focusing on this host-circuit coupling dynamics and gaining insightful results which in the long term help in designing robust and complex synthetic circuits also.  
\\ One of the most important reasons for this influence of cellular context is the dependency of the circuit on the host for the resources required for its gene expression. Here, by resource, we mean the cellular ingredients like RNAP, ribosome, ATP, protein degradation machinery \cite{borkowski2016overloaded,mcbride2017analyzing}, transcription factors, etc., that are supplied by the host cell for the gene expression process. 
Experiments prove that these resources are present in a limited manner inside cells \cite{Vind1993SynthesisOP,babu2003functional}. Recent experiments establish that the amount of available functional RNAP in the cell limits the transcription process majorly \cite{churchward1982transcription}. The effect of transcription factor sharing and its copy number in the gene expression process is shown in \cite{zabet2013effects} and it has been established that in an isogenic population of cells, this resource sharing enhances noise in the process of mRNA distribution \cite{das2017effect}.
These works indicate towards a major scope of re-exploring our well-known motifs for a resource-limited environment, and further emergent behaviors. 
To capture cellular resource sharing mathematically,  models have been developed by researchers considering different approaches (e.g., variable resource pool, resource availability as a function of cell growth rate etc.) \cite{qian2017resource,carbonell2016dealing,weisse2015mechanistic,darlington2018dynamic}.  Georgy et al. in their study, provide experimental evidence of cellular economy and proteins showing isocost-like expression while operating in a tight budget of ribosomes\cite{gyorgy2015isocost}. A recent study in this field reported major changes in toggle functionality as a consequence of resource competition \cite{chakraborty2021emergent,chakraborty2023Resource,chakraborty2021emergentb}. Theoretical models on competition of canonical and alternative sigma factors for RNAP in the steps of transcription initiation \cite{Grigorova} and transcription elongation \cite{Mauri2014AMF} show bacterial responses to environmental fluctuations. Several different approaches provide insightful results regarding changes of dynamical behavior because of different types of intracellular resource competition and emergent couplings due to this \cite{mcbride2017analyzing,9691638,goetz2022double,stone2022negatively}. 
\\ In this paper, we explore the emergence of richer dynamics in a three-gene motif, and comparing its response under resource limitation with a standard FFL architecture. 
Some recent experimental study with synthetic FFL systems have shown a stripe-like pattern for the third output gene node, in a collection of cells, with a varying concentration gradient of two input genetic nodes of FFL proteins \cite{santos2019using,schaerli2014unified,basu2005synthetic}. 
Keeping that in mind, in this work, we consider a motif that apparently, does not bear any resemblance with standard FFL, except for being a three-gene system. However, due to resource competition interesting responses can be observed, very similar to the FFL motif. Here, we show that the inherent structure of the FFL circuit is compensated by a limited pool of availability of translational resource (\textit{say, }ribosomes), while, resource competition plays a similar role to repression here. This is capable of establishing benefits of feed-forward control even in apparent absence of direct regulation. \\
In the upcoming sections, we report our findings as we perform a thorough spatiotemporal analysis of the three-gene motif under resource limitation and observe that it can behave like a conventional FFL motif. Unique responses of FFL motifs, like response delay, pulse formation, etc., are found in the temporal response of resource-driven motifs as well. We have further extended the model spatially and found the stripe patterns and the famous `Bull's eye' pattern \cite{santos2019using,schaerli2014unified} from the proposed resource-driven FFL in two dimensions. The paper is organized as follows: in Section. \ref{ffl concept} we have briefly discussed the conventional basic structure of FFL motifs, and, in an extension, the possibilities of occurrence of a feed-forward structure from the proposed three-gene architecture is also explained. In Section. \ref{model} we have discussed the model formulation of the proposed three gene motifs regulated by resource competition in detail. In Section. \ref{result}, we discuss the output responses giving rise to FFL-like behavior in transient scenarios as well as in a spatial multicellular diffusible environment. Finally, we conclude with some relevant discussion in Section. \ref{diss}.
\section{Feed-Forward Loops \& Resource Competition}\label{ffl concept}
\subsection{Conventional feed-forward loop motif}
In bacterial physiology, the presence of FFLs is found to modulate cellular dynamics very prominently. FFL motif is one of the most abundantly found motifs in nature where three genes having their unique pattern of regulation (activation or repression) give rise to coherent and incoherent motifs. In a three-gene motif (say $X$, $Y$ and $Z$), one regulating the next in series (i.e $X$ regulating $Y$, $Y$ is regulating $Z$), and also the first gene (say $X$) is regulating the third gene (say $Z$) in a direct fashion.  
Depending upon this mode of regulation (activation or repression) FFL motifs are conventionally classified into two groups, each containing four motifs, namely coherent FFL and incoherent FFL motifs. In the coherent type FFL motifs, the direct regulatory arm of $X \rightarrow Z$ is in harmony with the indirect regulation arm ($X$ regulating $Z$ via $Y$); these two are of opposite regulation in the case of incoherent motifs. 
The presence of further AND gate logic and OR gate logic specifies either the direct and indirect regulation in a  combined way regulates $Z$ production (AND logic) or any one of these regulations is sufficient to initiate $Z$ production (OR logic). In convention, an activator say $S_x$ and $S_y$ activates the proteins.
 Presence of activator, that is when $S_x=1$, the first protein is in active state $X$, and in absence of activator, $S_x=0$, implies $X=0$. For more precise undertsanding, the experimentally verified \textit{ara} system can be referred, where, $X$ = CRP, $Y$ = araC, $Z$ = araBAD, $S_x$ = cAMP and $S_y$ = L-arabinose \cite{mangan2003coherent}. 
\\ To proceed mathematically, the concentration of $Y$ and $Z$ can be represented by the set of equations, in a constitutive production of $X$:
\begin{equation*}
\frac{dY}{dt}=B_y+\beta_y\:\textit{f}(X,K_{xy})-Y\:\delta_y
\end{equation*}
\begin{equation}
\frac{dZ}{dt}=B_z+\beta_z\;\textit{G}(X,K_{xz},Y,K_{yz})-Z\:\delta_z\,\,,
\label{FFL}
\end{equation}
where, $\beta_i$, $(i \in \{y,z\})$ is the maximum production rate and $K_{ij}$ $(i,j \in \{x,y,z\},i \neq j)$ is the activation or repression coefficient, signifies the regulation by transcription factor $i$ on gene $j$. The AND gate function is represented by,
\begin{equation*}
    \textit{G}=\textit{f}(X,K_{xz})\;\;\textit{f}(Y,K_{yz})
\end{equation*}  
Here, the activator function is given by $\textit{f}(u,k)=\frac{(\frac{u}{k})^n}{1+(\frac{u}{k})^n}$ and repression function is represented by $\textit{f}(u,k)=\frac{1}{1+(\frac{u}{k})^n}$. $B_y$ and $\delta_y$ are the basal transcription rate and total degradation rate of $Y$ respectively, which includes the total dilution and degradation rates of $Y$ in a cell. $B_z$ and $\delta_z$ represents same for $Z$. $n$ is co-operativity which accounts for the multimer formation of proteins.

\subsection{Why resource-driven feed-forward mechanism can arise?}
Not all two gene input circuits act as FFL motifs in cells. For example, in \textit{E. coli} nearly $40\%$ of all existing two input operons function as FFL \cite{shen2002network}. In search of the reason for this, it is observed that the mutation in the binding sites of promoters can change the gene regulation, even to the extent of removing it completely, resulting in further non-existence of some regulatory linkages. These modifications in FFL may cause absence of $X$ to $Y$ regulation in some motifs, which are referred to as simple motifs or SM, (Fig. \ref{res motif}(c)) \cite{mangan2003structure}, and compared with conventional FFL motif to explain its unique regulatory behavior. Conventionally $X$ $\rightarrow$ $Z$ and $Y$ $\rightarrow$ $Z$ (say, $X$ regulates $Z$, and $Y$ regulates $Z$), these two regulations are considered essential to maintain AND gate regulation, and it can be said that the presence or absence of $X$ $\rightarrow$ $Y$ is the key factor that differentiates  FFL regulation and SM regulation. \\
Mutation in gene dynamics is sometimes biased by the preferences of bio-chemical reactions but majorly it is a random process \cite{loewe2008genetic}. It is possible that in some mutation the $X$ $\rightarrow$ $Z$ disappears leaving the rest of the wiring pattern intact. Essentially, this three-gene motif will no more behave like a conventional FFL. In this work, we have found that, in this scenario, if the two proteins develop a resource competition with asymmetric resource affinity, effective repression will come into the picture, which is strong enough to compensate for the effect of hill function type repression (of co-operativity 2) and the system shows similar response like FFL. In the next section, we have shown some three gene motifs where certain conventional direct regulations (similar to conventional FFL motifs like Fig. \ref{res motif}(a)), are not present, instead, some resource competition is generated due to limitations in the resource pool (as shown in Fig. \ref{res motif}(b)).  We have also considered the resource-driven simple regulation model in Fig. \ref{res motif}(d) which will be further used for evaluating the performance of the resource-driven FFL motifs. \\
For the rest of the text, for ease of discussion, the conventional feed-forward loop motifs,  proposed three gene motifs with resource competition, and resource-driven simple regulation model will be referred to as cFFL, rFFL, and rSM, respectively.

\section{Model Formulation}\label{model}
In our resource-sharing model, we focus especially on ribosome competition in the step of the translation process. To illustrate this process, we can take the example of \textit{yeast} where approximately 60000 mRNA molecule starts translating in parallel \cite{warner1999economics,zenklusen2008single}, while available ribosome (which is limited, nearly 240000 in \textit{yeast}) possibly scans the same transcript simultaneously. Now, if one mRNA starts accommodating ribosome with higher binding affinity, the other's translation initiation will be delayed (as the total supplier pool is getting affected) and suppressed as a result. Asymmetric binding affinity can have several reasons; the accessibility of the ribosome binding site on the mRNA significantly determines the basal translation level \cite{gualerzi2015initiation,cifuentes2019domains}, while Polycistronic mRNA pool contains a multiple ribosome binding sites (RBS) in most of the bacterial organisms \cite{burkhardt2017operon}. The recruitment of ribosomes to this RBS is temperature dependent. Temperature fluctuation induces re-folding of the mRNA which interacts with proteins and regulates the synthesis level in selective cases \cite{breaker2018riboswitches}. The nature of the environmental ligands also modulates this ribosome recruitment. \\
Considering this circumstance, let us consider a pathway where $X$ $\dashv$ $Y$ $\rightarrow$ $Z$ ($X$ represses $Y$, $Y$ activates $Z$). Suppose mRNAs for $X$ and $Z$ are involved in such a resource competition here, if $X$ recruits ribosomes more efficiently for its production, $Z$ will suffer a deficiency, resulting in repression in terms of resources. The motif is similar to cFFL coherent type 2, but here $X$ to $Z$ repression is not direct; instead, a resource competition serves the process. Before implementing mathematical equations, let us elaborate on our considerations and corresponding biological significance below:
\begin{figure}
    \centering
    \includegraphics[width=\textwidth]{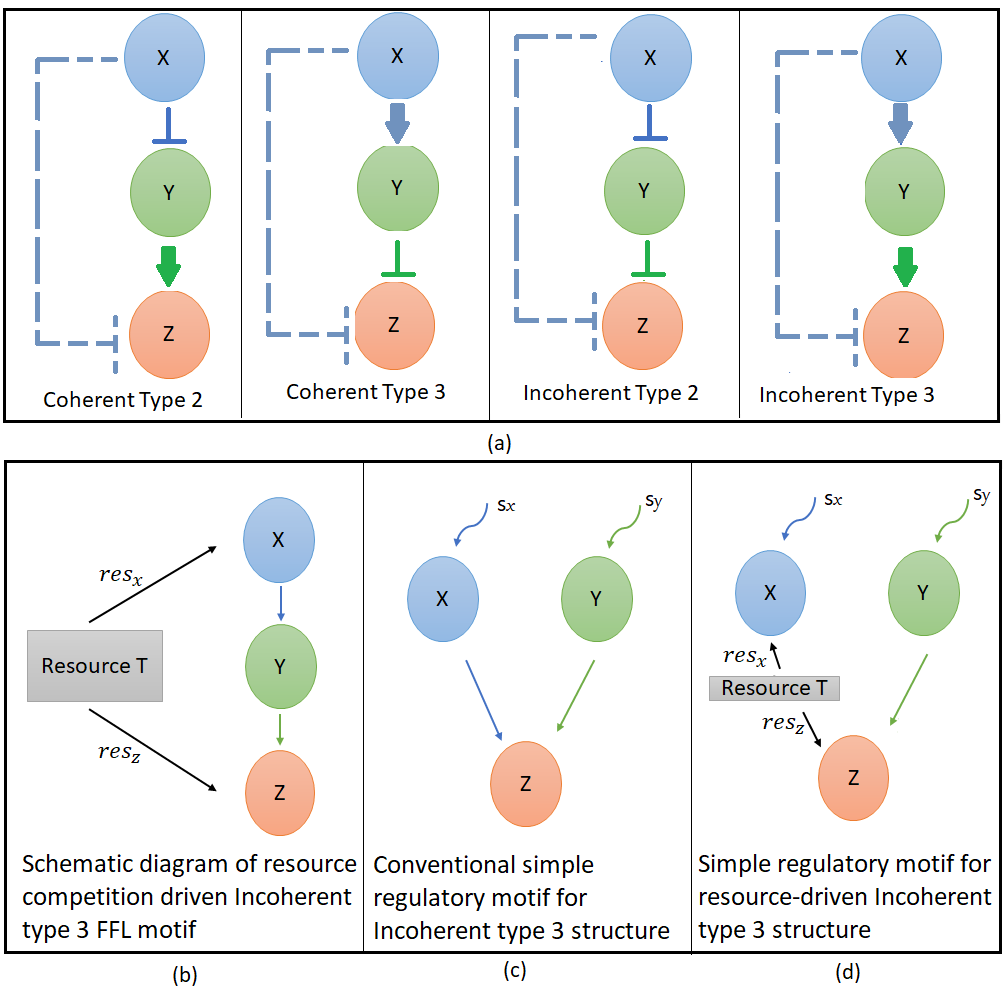}
    \caption{(a) Selected FFL motif among all motifs where $X$ to $Z$ regulation is of repression type. $X$ to $Z$ dotted line represents these repressions that can be replaced by resource competition for emergent resource-driven repression in these motifs. (b) Representing schematic diagram of proposed resource competition driven incoherent type 3 rFFL, structure. $X$ and $Z$ are collecting resource from the same pool $T$, with affinities respectively $res_x$ and $res_z$. (c) The SM motif of conventional incoherent type 3 FFL motif structure. $X$ and $Y$ are regulating $Z$, but there is no regulation of $X$ to $Y$. (d) Resource competition regulated simple motif structure, rSM, to study corresponding incoherent type 3 rFFL motif.}
    \label{res motif}
\end{figure}
\begin{itemize}
    \item We consider ribosomes to be distributed over several small cytoplasmic compartments in the cell. The limited presence of this translational resource in protein production is verified experimentally in some recent work \cite{gyorgy2015isocost}. We focus in the local resource pool here, present in the immediate vicinity of circuit of interest, which captures the circuit dynamics in a realistic way. Let $T$ represent the pool of ribosome, available for translation for its neighbourhood genes. 
    \item  The pool of  mRNA, as a result of transcription are respectively $g_x$ and $g_z$, which are ready to be translated into proteins $X$ and $Z$. Being expressed in the local field of cytoplasm, we consider both $X$ and $Z$ are collecting resource ribosomes from the same pool $T$.   
    \item The mRNA copies, which are ready for translation make a ribosome-bound complex in a step and get translated in the next step. The small sub-unit of ribosome binds to three initiation factors IF1, IF2, and IF3 along with a methionine-carrying tRNA first, then binds to mRNA and forms the complex. Let, $c_x$ and $c_z$ represent the ribosome bound complex of $X$ and $Z$ respectively. Now from the available total free pool $T$, $c_x$ and $c_z$ represents the bound complex, further free ribosome thus available for translation is given by ($T-c_x-c_z$).
    \item mRNA binds with ribosomes with a certain affinity. Let us consider $res_x$ and $res_z$ represents the resource affinity for $g_x$ and $g_z$ mRNA respectively. As discussed before, this affinity for resource allocation depends upon various factors. Thus resource affinity $res_x$ and $res_z$ can be taken as different taking care of all these biological factors \cite{gualerzi2015initiation,cifuentes2019domains,burkhardt2017operon,breaker2018riboswitches}.
    \item Protein is produced from the respective complex at a certain rate, $\epsilon_x$ and $\epsilon_z$ respectively for $X$ and $Z$.
    \item $\delta_x$, $\delta_y$, $\delta_z$, $\delta c_x$, $\delta c_z$ represents the overall degradation rates which account for the dilution and degradation inside the cell for respectively protein $X$, $Y$, $Z$ and complex $c_x$ and $c_z$.
    \item We achieve conventional AND gate logic of FFL, where both the direct regulation of $X$ to $Z$, and the indirect regulation on $Z$ via $Y$ acts combined as electronic AND gate logic, by multiplying $c_z \epsilon_z$ ($Z$ production term from its respective complex) with $Y$ regulatory term in our resource driven model.
    \item In \cite{mangan2003structure}, a step like behavior of $X$ induced by $S_x$ was considered. The same is achieved by allowing $X$ to produce from its complex for a time period say $t=0$ to $10$ when $S_x=1$. For our model, we consider $S_x=0$ blocks the complex formation $c_x$ at $t=10$, thus the protein $X$ is allowed to decay sharply, giving a nearly step-like production of $X$ wrt. time $t$.
    \item Respective rSM regulation model is represented by the same equation of $Z$ with $Y$ having constitutive expression, $Y=1$. As mentioned before a schematic diagram of the rSM regulation model is shown in Fig. \ref{res motif}(d).
\end{itemize}
Mathematical modeling for the type $2$ coherent FFL motif by our proposed resource competition model where $X$ to $Z$ repression arises due to resource competition is given by Eq. \ref{chr2}.
\begin{eqnarray}\label{chr2}
\frac{dc_x}{dt}&=&res_x\:(T-c_x-c_z)\:g_x-c_x\:\delta c_x\\
\frac{dc_z}{dt}&=&res_z\:(T-c_x-c_z)\:g_z-c_z\:\delta c_z \nonumber\\
\frac{dX}{dt}&=&c_x\:\epsilon_x-X\:\delta_x\nonumber\\
\frac{dY}{dt}&=&B_y+\beta_y\:\frac{1}{1+(\frac{X}{k_{xy}})^n}-Y\:\delta_y\nonumber\\
\frac{dZ}{dt}&=&B_z+\beta_z\:\frac{c_z\:\epsilon_z\:(\frac{Y}{k_{yz}})^n}{1+(\frac{Y}{k_{yz}})^n}-Z\:\delta_z\nonumber
\end{eqnarray}
Following similar arguments, we re-create resource-driven FFL (rFFL) capable of mimicking coherent type $3$, incoherent type $2$, and incoherent type $3$ rFFL motif. The reason behind choosing these four motifs is that in all these architectures $X$ to $Z$ regulation is a repression, which can also arise due to resource limitation.

\subsection{Model considerations for spatially extended dynamics of the three gene motif}
In order to study the spatiotemporal dynamics of rFFL motifs, we consider a two-dimensional cellular array, along two mutually perpendicular directions (say $a$ $\&$ $b$), considering a thin monolayer tissue of cells, each containing one motif. The proteins are allowed to diffuse beyond cell boundaries with particular diffusion rate constants, in a no-flux boundary condition. In two dimensions, for preliminary analysis, we have considered isotropic diffusion. Thus, $D_x$ represents the diffusion rate of protein $X$ along both the axes $a$ and $b$. Similarly, we define the diffusion coefficients as $D_y$ and $D_z$ for proteins $Y$ and $Z$ respectively. To achieve the concentration gradient of protein $X$, the corresponding mRNA pool $g_x$ is supposed to have a distribution profile mentioned accordingly in the result section (Eq. \ref{eqnsimple}) 
The concentration gradient of protein $Y$ is achieved by a distribution in protein expression profile $\beta_y$, mentioned in the result section, which will be discussed later (Eq. \ref{eqnsimple}) 
Here, for demonstration of experimentally reported patterns, we are elaborating on pattern formation by incoherent type 2 rFFL AND logic motif; other motifs can also be studied in a similar fashion. 

The representative equations for this in the presence of isotropic diffusion will be: 

\begin{eqnarray}\label{ichr22d}
\frac{dc_x}{dt}&=&res_x\:(T-c_x-c_z)\:g_x-c_x\:\delta c_x\\
\frac{dc_z}{dt}&=&res_z\:(T-c_x-c_z)\:g_z-c_z\:\delta c_z \nonumber\\
\frac{\partial X(a,b,t)}{\partial t}&=&c_x\:\epsilon_x-X\:\delta_x\;+\;D_x\frac{\partial^2X}{\partial a^2}+\;D_x\frac{\partial^2X}{\partial b^2}\nonumber\\
\frac{\partial Y(a,b,t)}{\partial t}&=&B_y+\beta_y\:\frac{1}{1+(\frac{X}{k_{xy}})^n}-Y\:\delta_y\;+\;D_y\;\frac{\partial^2Y}{\partial a^2}\;+\;D_y\frac{\partial^2Y}{\partial b^2}\nonumber\\
\frac{\partial Z(a,b,t)}{\partial t}&=&B_z+\beta_z\:\frac{c_z\:\epsilon_z}{1+(\frac{Y}{k_{yz}})^n}-Z\:\delta_z\;+\;D_z\;\frac{\partial^2Z}{\partial a^2}\;+D_z\frac{\partial^2Z}{\partial b^2}\nonumber
\end{eqnarray}

\section{Results}\label{result} 
We have investigated resource-driven FFL structures, rFFL, similar to Fig. \ref{res motif}(b), in comparison to resource-driven simple motifs, rSM (similar to Fig. \ref{res motif}(d)), for different resource allocation rates. Instead of the topological differences, we find that rFFL motifs show similar responses as cFFL motifs (shown in Fig . \ref{res motif}(a)), as an emergent response of resource competition. 
\subsection{Transient response of rFFL motifs:}
\subsubsection{Coherent type 2 rFFL motif:}\label{c2}
\begin{figure}
    \centering
    \includegraphics[width=\textwidth]{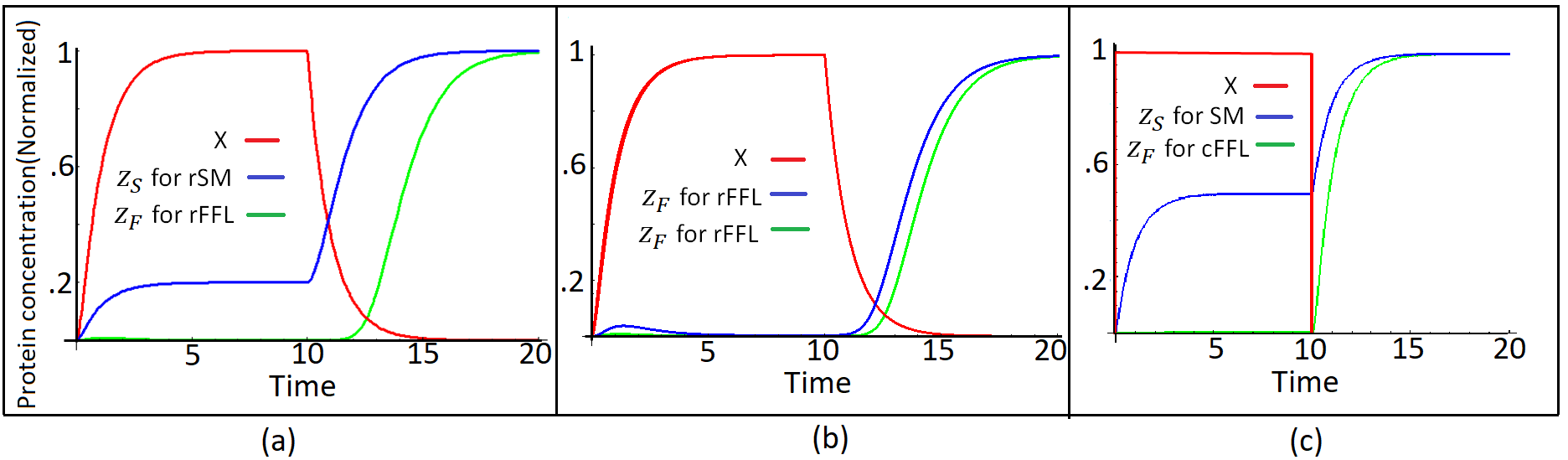}
    \caption{Comparative analysis of Kinetic behavior of coherent type 2 rFFL motif and cFFL motif in AND gate logic. (a). The behavior of coherent type 2 rFFL motif. The red line represents $X$. The green and the blue line represents $Z_F$ in rFFL and $Z_S$ in rSM motifs, respectively. Note that wrt. the blue line (the rSM logic), the green curve (rFFL logic) shows a delay in reaching the steady state. (b). Variation in resource affinity affects the delay in reaching the steady state of $Z_F$ in rFFL. The red line is for $X$. Blue line is for $Z_F$, $res_x = 1, res_z = 1$ and Green line is for $Z_F$, when $res_x = 1, res_z = 0.05$ in Coherent type 2 rFFL. (c). Kinetics of Coherent type $2$ cFFL AND gate motif \cite{mangan2003structure}. $k_{xy} = 0.1, k_{yz} = 0.1, k_{xz} =1, n = 2$. The red line is for $X$, the Green line shows the nature of $Z_F$ in cFFL motif response, and the blue line is for $Z_S$, for the SM motif.    $\beta_y = \beta_z= 1$, $\delta_x = \delta_y = \delta_z = 1, \delta c_x = 1, \delta c_z = 1, \epsilon_x = 1, \epsilon_z = 1, g_x = 5, g_z = 5, T=10, B_y = B_z =0$ for both (a) and (b).}
    \label{chrnt2}
\end{figure}

The gene circuits we are going to compare here are the following:
\begin{itemize}
    \item Coherent type 2 rFFL motif with coherent type 2 cFFL motif
    \item Coherent type 2 rFFL motif with corresponding rSM motif.
\end{itemize}
The output notation will be considered as $Z=Z_F$ in the case of the FFL structure, while the output notation will be considered as $Z=Z_S$, for the simple structure. The initial results are reported in Fig. \ref{chrnt2}(a), where the green and the blue line represent $Z_F$ and $Z_S$ for rFFL output and rSM output respectively, while the red line indicates input $X$. We analyze the response of output protein upon step-like addition of inducer $S_x$, in the presence of $Y$ in AND logic rFFL coherent type $2$ motif. At the time $t=0$, $g_x$ starts translation, producing a complex $c_x$ to further synthesize $X$, by collecting resource ribosome with affinity $res_x$ from the pool $T$. At the same time, $g_z$ also started to produce its complex $c_z$, and thus allocated resources at a rate of $res_z$. Considering the higher value of $res_x$ ($res_x> res_z$), an effective repression on $Z$ emerges, giving rise to rFFL architecture, as $X$ gets produced at the cost of $Z$. 
If we observe $Z_F$ in Fig. \ref{chrnt2}(a) and (c), rFFL response can be compared with the corresponding cFFL. Due to resource-driven effective repression of $Z$ by $X$, and direct repression $X$ $\dashv$ $Y$ $\rightarrow$ $Z$,  $Z$ is low throughout the active $X$ state; this dynamics is same as cFFL \cite{mangan2003structure}. Moreover, similar to the cFFL motifs, in our rFFL motif, $Z$ goes on with $X$ off. This is a demonstration of the well-known inverted output of coherent type 2 architecture, as observed in rFFL here. We also point out that the steady state logic of $Z$ in our rFFL motif is both sensitive to $S_x$ and $S_y$, similar to cFFL motif. Putting $Y$ in the off state by $S_y = 0$, $Z$ steady state goes off (not shown here).\\
Now, to compare rFFL with the corresponding simple architecture, rSM, we follow a similar convention as \cite{mangan2003structure}. At $t=10$,  we make $S_x=0$, while $S_x$ is the inducer of $X$ production. The red line of $X$ in Fig. \ref{chrnt2} thus shows the behavior of protein production kinetics with respect to $S_x$ on step at $t = 0$, and $S_x$ off state at $t = 10$. This blocks the complex of $X$ production, $c_x = 0$. As soon as $c_x$ production stops, $X$ drastically falls to zero and we investigate the dynamics of $Z$ here. 
With respect to the resource-driven simple motif, rSM (Fig. \ref{res motif}(d)), where $X$ to $Z$ resource competition and $Y$ to $Z$ regulation works separately, we find the rFFL-like structure shows a delay in reaching $Z$ steady state as shown in Fig. \ref{chrnt2}(a). The response time of protein $Z$, defined as the time to reach $50\%$ of its final concentration \cite{rosenfeld2003response,savageau1976biochemical} is greater in the rFFL motif than that of the rSM motif. This is a signature response of coherent type 2 cFFL, and shown in Fig. \ref{chrnt2}(c). 
\\Finally, we study the effect of the variation in resource allocation rates. This determines the efficiency of collecting resources for the production of the protein and significantly modulates the delay here as shown in Fig. \ref{chrnt2}(b).  $res_x = 1, res_z = 0.05$ the green representative of protein $Z_F$ is delayed wrt. $res_x = res_z =1$ the blue line. Theoretically, variation in resource allocation rate tunes the strength of repression on $Z_F$, thus, delay in $Z_F$ synthesis accordingly gets modulated. 

\subsubsection{Key characteristic features of conventional FFL demonstrated by rFFL:}
\begin{figure}
    \centering
    \includegraphics[width=\textwidth]{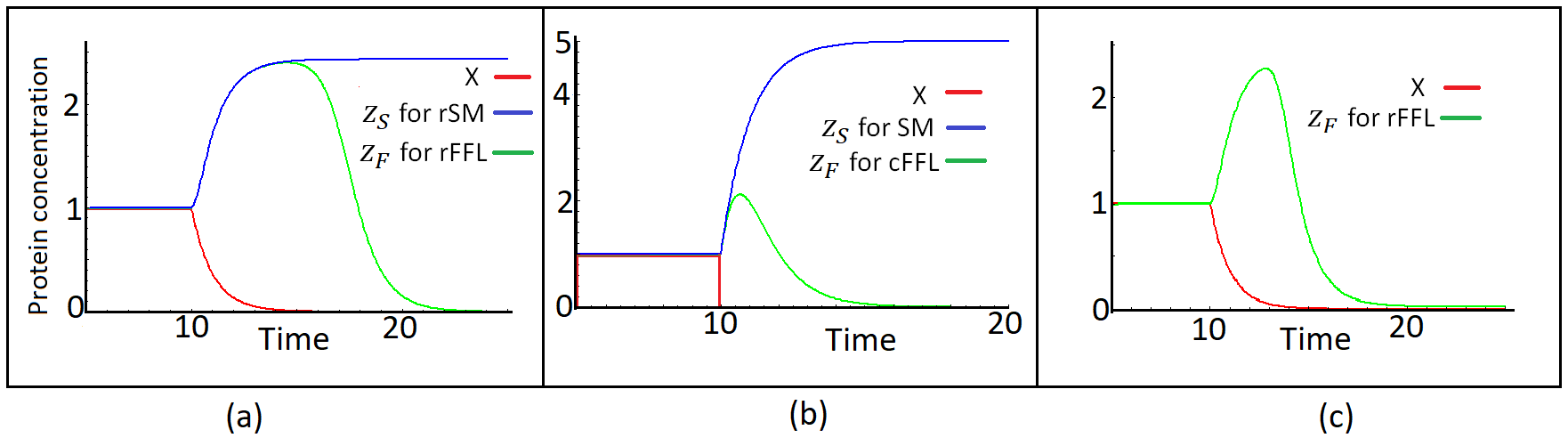}
    \caption{Pulse formation in AND logic rFFL incoherent motifs and comparative cFFL motif output response. (a) Resource-driven incoherent type $3$ motif. $ k_{xy} = 0.1, k_{yz} = 0.1, res_x = 1, res_z = 0.5$. A pulse is seen in the $S_x$ off state for $Z_F$ response. The blue is for $Z_S$ in the rSM regulation system, and the green is for $Z_F$ in the rFFL motif response output (b). Incoherent type $3$ cFFL motif. $ k_{xy} = 1, k_{yz} = 0.5, k_{xz} = 0.5$. The blue is for $Z_S$ in the SM regulation system, and the green is for $Z_F$ in the cFFL motif response output (c). Pulse formation in incoherent type $2$ rFFL motif. The blue is for $Z_S$ in the rSM system, and the green is for $Z_F$ in the rFFL motif response output.Parameter values are $k_{xy} = 0.1, k_{yz} = 0.1, res_x = 1, res_z = 0.5$. For both (a) and (c) $n = 2, T = 10, \delta c_x = 1, \delta c_z = 1, \epsilon_x = 1, \epsilon_z = 1, g_x = 5, g_z = 5, B_y = B_z = 0, \beta_y = \beta_z = 1$.}
    \label{pulse}
\end{figure}

Following the same approach as described in \ref{c2}, we explore other rFFL architectures as shown in Fig. \ref{res motif}(a). We have consolidated the major observations and findings below:
\begin{itemize}

\item{\textit{\textbf{$S_x$ off state delay in coherent rFFL motif:}}}\\
 $Z_F$ steady state is delayed in coherent type $2$ and type $3$ rFFL motif wrt. $Z_S$ of rSM motif in $S_x$ off state.  The behavior is similar to the conventional motif responses \cite{mangan2003structure}. Additionally, we have also observed that the variation in resource affinity value regulates the delay response significantly (as shown in Fig. \ref{chrnt2}(b), Fig. \ref{chrnt3}(b)). \\

\item{\textit{\textbf{Steady state logic of $Z_F$ is dependent on $S_x$ in coherent  rFFL:}}}\\
Steady-state logic of $Z_F$ is inverted with $S_x$ in coherent type $2$ and coherent type $3$ rFFL motif. That is, $Z_F$ in rFFL goes on in $S_x$ off step. As mentioned earlier, $S_x$ off state releases the repression on $Z_F$, both in terms of resource and indirect regulatory repression via $Y$, in both coherent type $2$ and $3$ rFFL motif, thus, $Z_F$ production increases, making $Z_F$ response inverted with $S_x$ in rFFL. \\
\begin{figure}
    \centering
    \includegraphics[width=\textwidth]{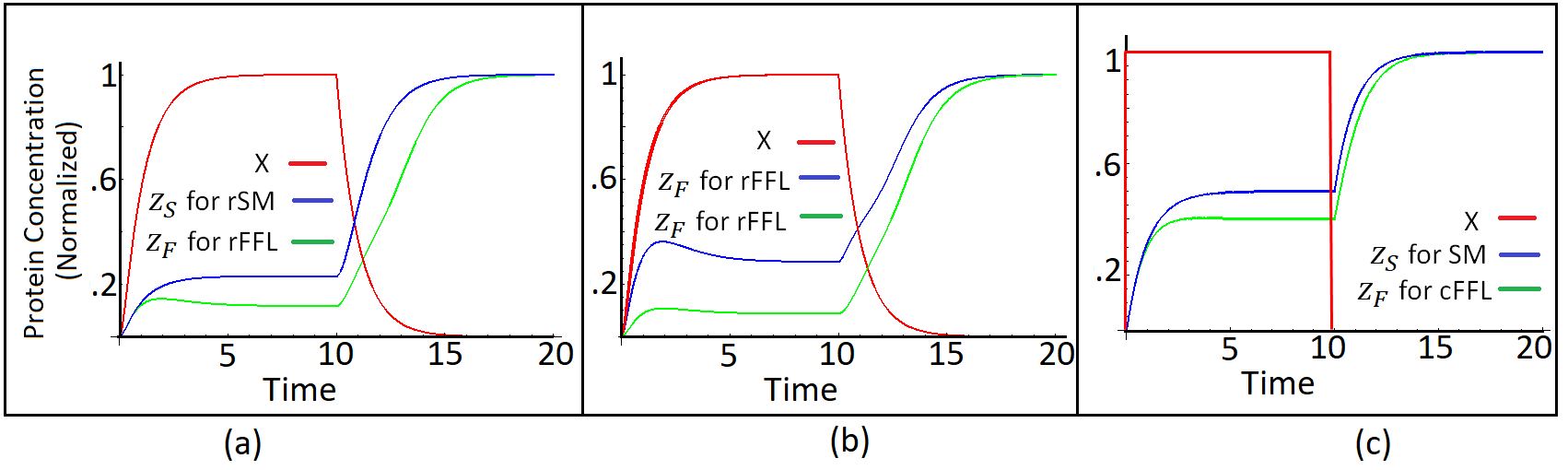}
    \caption{Comparative analysis of Kinetic behavior of coherent type 3 rFFL motif and cFFL motif in AND gate logic. (a). Kinetic behavior of Coherent type 3 rFFL motif. The red line represents $X$, the green line is for rFFL $Z_F$ response, and the blue line is $Z_S$ for rSM logic. Note that wrt. the blue line (the rSM logic) the green curve (the rFFL logic) show a delay in reaching the steady state. (b). Variation in resource affinity affects the delay in reaching the steady state of $Z_F$ in rFFL. The red line is for $X$. Blue line is for $Z_F$, $res_x = 1, res_z = 1$. Green line is for $Z_F$, when $res_x = 1, res_z = 0.01$. (c). Kinetics of coherent type $3$ cFFL AND gate motif \cite{mangan2003structure}. The red line is for $X$, the Green line shows the nature of $Z_F$ in the cFFL motif response, and the blue line is for $Z_S$ in the SM motif. Parameter values are $k_{xy} = k_{yz} = k_{xz} = 1, n = 2,$. For (a), (b) and (c) rest parameters are  $\beta_y = \beta_z= 1$, $\delta_x = \delta_y = \delta_z = 1$, $B_y = B_z = 0$.}
    \label{chrnt3}
\end{figure}
\item{\textit{\textbf{Steady state of $Z_F$ is dependent on $S_y$ in coherent type 2 rFFL motif but not in coherent type 3 rFFL :}}}\\
The steady state of $Z_F$ responds strongly to $S_y$ in the case of coherent type $2$ rFFL, and the inverted output nature is lost when $S_y=0$ (at $S_x = 0$, $S_y = 1$ the $Z_F$ steady state is inverted in nature). But the steady state of $Z_F$ is not dependent on $S_y$ in the case of coherent type $3$ rFFL motif. Both these behaviors are similar to the cFFL motif responses. \\

\item{\textit{\textbf{Pulse generation in $S_x$ off state of incoherent rFFL motifs:}}}\\
Incoherent type $2$ and $3$ cFFL model motif shows pulse formation in $S_x$ off state for AND gate logic. Our AND logic rFFL model shows similar results in output as shown in Fig. \ref{pulse}(a) and \ref{pulse}(c). $X$ and $Z$ are collecting resource from the pool $T$, and thus $X$ is putting a repression in $Z$ production, strength depending upon resource affinity. Now, for incoherent type $3$ motif, at $t=10$, the complex formation of $X$, that is, in a straightforward way, the production of $X$ is blocked, and no resource demands $X$ are valid as well. Thus, the available resource pool is now open for $Z$, and the repression in terms of resources is no more. Thus, $Z$ production suddenly increases at $S_x$ off state. The wiring architecture shows $X$ activates $Y$, which is also an activator of $Z$. Now when $X=0$, this $X$ can't activate $Y$, eventually, $Y$ production decreases, and further $Z$ production decreases, as $Y$ is not produced enough so it can't activate $Z$. Thus, $Z$  (or more specifically $Z_F$ for rFFL incoherent type 3) decays eventually. Thus, $Z_F$ for this rFFL incoherent type 3 motif shows a pulse in output, as the sudden increase in production eventually dies out. Similarly, the pulse formation in the incoherent type $2$ rFFL motif can be explained.\\ 

\item{\textit{\textbf{No pulse is created in $S_x$ on state of incoherent rFFL motifs:}}}\\
The incoherent type $2$ and $3$ cFFL do not generate a pulse in response to $S_x$ on the step. The rFFL motifs show similar results (Fig. \ref{pulse}(a) and \ref{pulse}(c)).\\

\item{\textit{\textbf{Steady state behavior of incoherent rFFL with no basal activity, $S_y$ effect:}}}\\
The steady-state logic of the Incoherent type 2 rFFL motif is found to depend on $S_y$. In the presence of $S_y$, $Z_F$ creates a pulse and then comes down to a low state eventually, while in the absence of $S_y$, the steady state is high, and no pulse is created. But type $3$ incoherent rFFL motif has a constant steady state $0$, which does not depend upon $S_y$. These behaviors are similar to the cFFL behaviors as well. \\
\item{\textbf{\textit{$S_y$ effect in pulse generation in rFFL incoherent motifs:}}}\\
Similar to the cFFL motif, in our rFFL motif, $Z_F$ shows a pulse in output when $S_y$ is on, but $Z_F$ is high and steady when $S_y$ is off. In incoherent type $3$ rFFL motif, $Z_F$ shows no pulse in $S_y$ off state.
\end{itemize}
\subsection{Reaction diffusion model: Pattern formation by incoherent type 2 rFFL motif}
In a tunable concentration gradient of the two input gene circuits, the third output gene of the cFFL structure is found to produce protein patterns. A stripe pattern is achieved because of the topological structure of cFFL motifs. In incoherent type 2 cFFL topology, the third gene is only active for the intermediate concentration of two input genes, forming a Bandpass filter in output. Here, we must mention the `Bull's eye' pattern also, in a controlled concentration gradient of the first two genes, the third output gene shows the `Bull's eye' pattern in a synthetic experimental environment.  In 2005, Basu et al. first synthetically displayed a stripe pattern in a population of \textit{E. coli} in a  \textit{Vibrio fischeri} quorum sensing system by using incoherent type 2 topology \cite{basu2005synthetic}. In a morphogen analog, the acyl-homoserine lactone (AHL) signal is produced by a localized source of `sender' cells and diffuses across an agar plate. AHL gradient is interpreted by uniformly distributed `receiver' cells, which result in a low-high-low pattern of a fluorescent reporter. Several synthetic experiments successfully recreated the Bull's eye and stripe patterns by the ``\textit{favorite}" incoherent type 2 topology \cite{santos2019using,schaerli2014unified,basu2005synthetic,ellis2009diversity} and incoherent type 3 topology \cite{greber2010engineered,saxena2016programmable,kampf2012rewiring}. 
In further advancement, multiple stripe-forming networks are connected together and exposed to two morphogen gradients in order to create more complex patterns such as cross patterns. We find similar results with our proposed rFFL motifs. Instead of lacking the exact FFL structure, the proposed resource competition fulfills the hill function driven $X$ to $Z$ repression of cooperativity $2$ and similar patterns we have recreated in a multicellular environment. We have incorporated the effect of diffusion into the system. Simulation and further mathematical analysis refer to the high stability of the patterns in a long time limit.  
\\Here, we have considered a two-dimensional array of $(200\times 200)$ cells. The position of each cell is discretized in direction $a$ as $a_i$, where $i \in ( 1,200 )$ and in $b$ direction as $b_j$ where $j\in(1,200)$.

\subsubsection{Spatiotemporal pattern formation by rFFL motif:} 
\begin{figure}
    \centering
    \includegraphics[width=\textwidth]{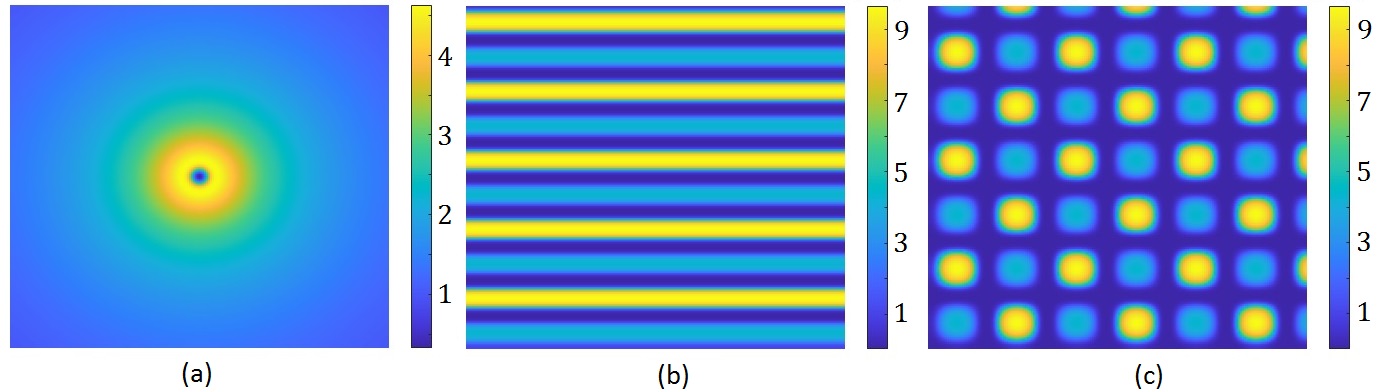}
    \caption{Formation of elementary spatiotemporal patterns via rFFL incoherent type 2 motif. (a) Bull's eye pattern in rFFL motif. (b) Stripe pattern in rFFL motif. (c) Block pattern in rFFL. 
 The parameter value for (a) $k_1=0.15,\;k_2=100, k_3=25$. (b) $k_1=1, \;k_2=20, \; k_3=25$. (c) $k_1=1,\;k_2=10, \;k_3=25$ The diffusion coefficient for all the figures is $D_x\;=D_y\;=D_z=0.5$. The rest of the parameter values $k_{xy}=k_{yz}=0.1, res_x=1, res_z=0.5,\epsilon_x=1,\beta_z=1, T=10, n=2, \delta c_x=\delta c_z= \delta_x =\delta_y=\delta_z=1, \epsilon_z=1, B_y=B_z=0.$  }
    \label{simple}
\end{figure}

In a two-dimensional arrangement of the cellular arena, each containing one motif of resource-driven incoherent type 2 AND logic feed-forward structure, we find spatiotemporal patterns of protein $Z$ in the output. We have achieved the \textit{famous} `Bull's eye' pattern of the cFFL structure in our rFFL structure via the initial condition of Eq. \ref{eqnsimple}(a). The corresponding `Bull's eye' pattern in our rFFL incoherent type 2 motif is shown in Fig. \ref{simple}(a). Moreover, initial conditions giving rise to periodic structures (periodic in one of the tqo dimensions, resulting into \textit{stripes} and periodic in both dimensions, resulting into \textit{blocks}) is common in biological systems. Thus, setting initial conditions as Eq. \ref{eqnsimple}(b), (c) we have received the stripe pattern (Fig. \ref{simple}(b)) and further a block pattern in Fig. \ref{simple} (c). 
\begin{subequations}\label{eqnsimple}
\begin{align}
 g_x(a_i,b_j,0)&=k_1\;\sqrt{(i-k_2)^2+(j-k_2)^2}&  \beta_y(a_i,b_j,0)&=k_3   \\
g_x(a_i,b_j,0)&=k_1\;\sin({\frac{\pi \; i}{k_2}})&  \beta_y(a_i,b_j,0)&=k_3    \\
g_x(a_i,b_j,0)&=k_1\;\sin({\frac{i}{k_2}})\;\sin{(\frac{j}{k_2})}&  \beta_y(a_i,b_j,0)&=k_3   
\end{align}
\end{subequations}
The patterns are stable through the long simulation time and further for a wide range of diffusion coefficients (say $D_x\;=D_y=\; D_z=0.01$ to $D_x\;=D_y=\; D_z=1.5$ and even beyond this)

\subsection{Stability of the rFFL system}
\subsubsection{Temporal Stability}
The last three equations in Eq. \ref{ichr22d} are nonlinear coupled ordinary differential equations in the form of the Reaction equations, where we rename the Reaction terms in the RHS as the following:
\begin{eqnarray}
f(X, Y, Z) &=& c_x \epsilon_x - X \delta_x \nonumber \\
g(X, Y, Z) &=& B_y + \beta_y \frac{1}{1+(\frac{X}{k_{xy}})^n} - Y \delta_y \nonumber \\
h(X, Y, Z) &=& B_z + \beta_z \frac{c_z \epsilon_z}{1+(\frac{Y}{k_{yz}})^n} - Z \delta_z
\end{eqnarray}
To analyze the stability of the spatiotemporal dynamics of the system, and for the analytical tractability, we linearize (up to first order) this inherently nonlinear system at its steady state $S_* = (X_0, Y_0, Z_0)$. Thus we have constructed a homogeneous steady state $S_*$, stable under diffusion. The linearization gives
\begin{eqnarray}
f(X, Y, Z) & \approx & f(X_0, Y_0, Z_0) + f_X \Delta_X + f_Y \Delta_Y + f_Z \Delta_Z \nonumber \\
g(X, Y, Z) & \approx & g(X_0, Y_0, Z_0) + g_X \Delta_X + g_Y \Delta_Y + g_Z \Delta_Z \nonumber \\
h(X, Y, Z) & \approx & h(X_0, Y_0, Z_0) + h_X \Delta_X + h_Y \Delta_Y + h_Z \Delta_Z
\end{eqnarray}
where $f_{\eta} $, $g_{\eta} $, and $h_{\eta}$, $(\eta=X, Y, Z)$ are the first derivatives of the Reaction Terms calculated at the steady state $S_*$ as a function of the concentrations of the three proteins $X$, $Y$, and $Z$. The explicit expressions in terms of other parameters of the system are as follows
\begin{eqnarray}
f_X & = & \frac{\partial f}{\partial X} \bigg|_{(X_0, Y_0, Z_0)} = -\delta_x \nonumber \\
f_Y & = & \frac{\partial f}{\partial Y} \bigg|_{(X_0, Y_0, Z_0)} = 0 \nonumber \\
f_Z & = & \frac{\partial f}{\partial Z} \bigg|_{(X_0, Y_0, Z_0)} = 0 \nonumber \\ \\
g_X & = & \frac{\partial g}{\partial X} \bigg|_{(X_0, Y_0, Z_0)} = -\frac{n \beta_{y} (\frac{X_0}{k_{xy}})^{n-1}}{k_{xy} \big(1+ (\frac{X_0}{k_{xy}})^n \big)^2 } \nonumber \\
g_Y & = & \frac{\partial g}{\partial Y} \bigg|_{(X_0, Y_0, Z_0)} = - \delta_y \nonumber \\
g_Z & = & \frac{\partial g}{\partial Z} \bigg|_{(X_0, Y_0, Z_0)} = 0 \nonumber \\ \\
h_X & = & \frac{\partial h}{\partial X} \bigg|_{(X_0, Y_0, Z_0)} = 0 \nonumber \\
h_Y & = & \frac{\partial h}{\partial Y} \bigg|_{(X_0, Y_0, Z_0)} = -\frac{n c_z \epsilon_z \beta_{z} (\frac{Y_0}{k_{yz}})^{n-1}}{k_{yz} \big(1+ (\frac{Y_0}{k_{xy}})^n \big)^2 } \nonumber \\
h_Z & = & \frac{\partial h}{\partial Z} \bigg|_{(X_0, Y_0, Z_0)} = -\delta_z \; .
\end{eqnarray}
And $\Delta{\eta}$ are the perturbations from the steady state $S_*$, given by $\Delta{\eta} = {\eta}(t) - {\eta}_0 $.
Thus the linearized system
 \begin{equation} \label{eq:stabilityMatrix}
\frac{\partial \Delta \eta}{\partial t}
= \mathcal{A}  \Delta \eta
\end{equation}
where $\mathcal{A}$ is the Linear Stability Matrix
\[
\mathcal{A} =
\begin{bmatrix}
f_{X} & f_{Y} & f_{Z} \\
g_{X} & g_{Y} & g_{Z} \\
h_{X} & h_{Y} & h_{Z}
\end{bmatrix}
\]
and
\[  \Delta \eta =
\begin{bmatrix}
    \Delta X \\
    \Delta Y \\
    \Delta Z
\end{bmatrix} \;
\]
is the perturbation matrix.
The stability of the system can be determined by the signs of the trace $\tau$, determinant $\mathcal{D}$, and the eigenvalues $\lambda_{i}$s of the linear stability matrix $\mathcal{A}$ of the system. For the system we consider here, the quantities are given by the following
\begin{eqnarray} \label{eq:TauDet}
\tau &=& Tr(\mathcal{A}) = f_X + g_Y + h_Z \nonumber \\
&=& - (\delta_x + \delta_y + \delta_z ) < 0 \nonumber \\
\mathcal{D} &=& Det(\mathcal{A}) = f_X (g_Y h_Z - g_Z h_Y) + f_Y (g_Z h_X - g_X h_Z) + f_Z (g_X h_Y - g_Y h_X) \nonumber \\
&=& - \delta_x \delta_y \delta_z < 0 \nonumber \\
\lambda_i &=& ( - \delta_x, - \delta_y, - \delta_z ) < 0
\end{eqnarray}
We have considered $\delta_x$, $\delta_y$, and $\delta_z$ to be positive real numbers ($\delta_x = \delta_y = \delta_z = 1$). The negative trace, determinant, and eigenvalues for a three-variable system signify that the system is stable under temporal perturbation. 

\subsubsection{Spatiotemporal Stability}
In a Reaction system i.e. a well-mixed system (without diffusion), there is a spatially uniform steady state which remains stable to perturbations. Whereas in a Reaction-Diffusion system i.e. a non-mixed system, this steady state becomes unstable because of diffusion. Usually, diffusion is a stabilizing process that homogenizes the system. But in our case, from the interactions of three stabilizing processes, diffusion-driven instability or spatiotemporal patterns (as in Fig. \ref{simple}) spontaneously emerge.

We assume that the system has a stable stationary solution in the absence of molecular diffusion of the proteins. Spatiotemporal stationarity requires
\begin{eqnarray}
\frac{\partial X}{\partial t} = 0 , \;\;\;\;  \frac{\partial Y}{\partial t} = 0 , \;\;\;\; \frac{\partial Z}{\partial t} = 0 \nonumber \\
\frac{\partial^2 X}{\partial a^2}=0 , \;\;\;\; \frac{\partial^2 Y}{\partial a^2}=0 , \;\;\;\; \frac{\partial^2 Z}{\partial a^2}=0 \nonumber \\
\frac{\partial^2 X}{\partial b^2}=0 , \;\;\;\; \frac{\partial^2 Y}{\partial b^2}=0 , \;\;\;\; \frac{\partial^2 Z}{\partial b^2}=0
\end{eqnarray}
which in turn implies
\begin{equation}
f(X,Y,Z) = 0 , \;\;\;\; g(X,Y,Z) = 0 , \;\;\;\; h(X,Y,Z) = 0
\end{equation}
Linearizing around the stationary state $S_*$ and setting $S_* = (X, Y, Z) = (X_0, Y_0, Z_0) + (\Delta X (a, b, t), \Delta Y (a, b, t), \Delta Z (a, b, t))$, or $S_* = \eta_0 + \Delta \eta (a,b,t)$ in short, we obtain the following linearized system,
\[ \label{}
\frac{\partial}{\partial t} \Delta \eta
=
\mathcal{A}  \Delta \eta
+ \mathbf{D} \bigg( \frac{\partial^2}{\partial a^2} + \frac{\partial^2}{\partial b^2} \bigg)  \Delta \eta
\]
$\mathcal{A}$ and $\Delta \eta$, the linear stability and perturbation matrices, are the same as given in Eq.\ref{eq:stabilityMatrix}. The diffusion matrix $\mathbf{D}$ takes the following form
\[
\mathbf{D}= \begin{bmatrix}
    D_X  & 0 & 0 \\
    0 & D_Y & 0 \\
    0 & 0 & D_Z \\
\end{bmatrix}
\]
The Fourier Transform of the equation
\[ \label{}
\frac{\partial}{\partial t} \widehat{\Delta \eta}
=
(\mathcal{A} 
-2 k^2 \mathbf{D}) \widehat{\Delta \eta}
\]
where $\mathbf{k} = (k_a, k_b) $ is the vector of wavenumbers in two dimensional space and $\widehat{\Delta \eta} = 
\begin{bmatrix}
    \widehat{\Delta X} \\
    \widehat{\Delta Y} \\
    \widehat{\Delta Z}
\end{bmatrix}$ is the matrix of the Fourier-transformed perturbations. This is now a set of linear ordinary differential equations in the Fourier-transformed variables $\widehat{\Delta X}$, $\widehat{\Delta Y}$, $\widehat{\Delta Z}$ for the protein concentrations. The factor $2$ comes from consideration of both the spatial directions $a$ and $b$. We note that solving the above linear system is the same as searching for harmonic solutions to equations to the linearized partial differential equations
\begin{eqnarray}
    \Delta X &=& \Delta X_0 \exp{-(i\Vec{k}.\Vec{a}+j\Vec{k}.\Vec{b})+\lambda t} \nonumber \\
    \Delta Y &=& \Delta Y_0 \exp{-(i\Vec{k}.\Vec{a}+j\Vec{k}.\Vec{b})+\lambda t} \nonumber \\
    \Delta Z &=& \Delta Z_0 \exp{-(i\Vec{k}.\Vec{a}+j\Vec{k}.\Vec{b})+\lambda t}
\end{eqnarray}
Substituting these and simplifying,
\[
\lambda \widehat{\Delta \eta}
= \mathcal{S_{\mathcal{D}}}
\widehat{\Delta \eta}
\]
where $\mathcal{A_{\mathcal{D}}} = \mathcal{A}-2 k^2 D$ is the Linear Stability Matrix of the Reaction-Diffusion system. In the presence of diffusion, the Trace of $\mathcal{S_{\mathcal{D}}}$ is given by
\begin{eqnarray}
    \tau_{\mathcal{D}} &=& f_X + g_Y + h_Z - k^2 (D_X+D_Y+D_Z) \nonumber \\
    &=& - (\delta_x + \delta_y + \delta_z ) - 6 k^2 D
\end{eqnarray}
We note here that the factor $6$ comes from the consideration of the two spatial dimensions $a$ and $b$, and three protein concentrations $X$, $Y$, and $Z$, and, for simplicity, we consider the diffusion coefficients to be the same for all proteins in both the spatial directions.
The determinant of $\mathcal{S_{\mathcal{D}}}$ is given by
\begin{eqnarray}
    \mathcal{D}_D &=& f_Y (g_Z h_X - g_X h_Z + 2 D g_X k^2) + 
 f_Z (-g_Y h_X + g_X h_Y + 2 D h_Y k^2) \nonumber \\
 && + (f_X - 2 D k^2) (-g_Z h_Y + (g_Y - 2 D k^2) (h_Z - 2 D k^2)) \nonumber \\
    &=& - (\delta_x + 2 D k^2) (\delta_y + 2 D k^2) (\delta_z + 2 D k^2)
\end{eqnarray}
and the eigenvalues
\begin{equation}\label{lambda}
   \lambda_i^D = (-\delta_x - 2 D k^2, - \delta_y - 2 D k^2, -\delta_z - 2 D k^2)
\end{equation}
Since $\delta_{\eta}$, $k$, and $D$ are positive and real, the trace $\tau_D$, determinant $\mathcal{D}_D$, and the eigenvalues $\lambda_i^D$ of the Reaction-Diffusion system are negative, suggesting the system to remain stable under spatiotemporal perturbations, or in other words, during the interplay of both Reaction and Diffusion.

\begin{figure}
    \centering
    \includegraphics[width=0.75\textwidth]{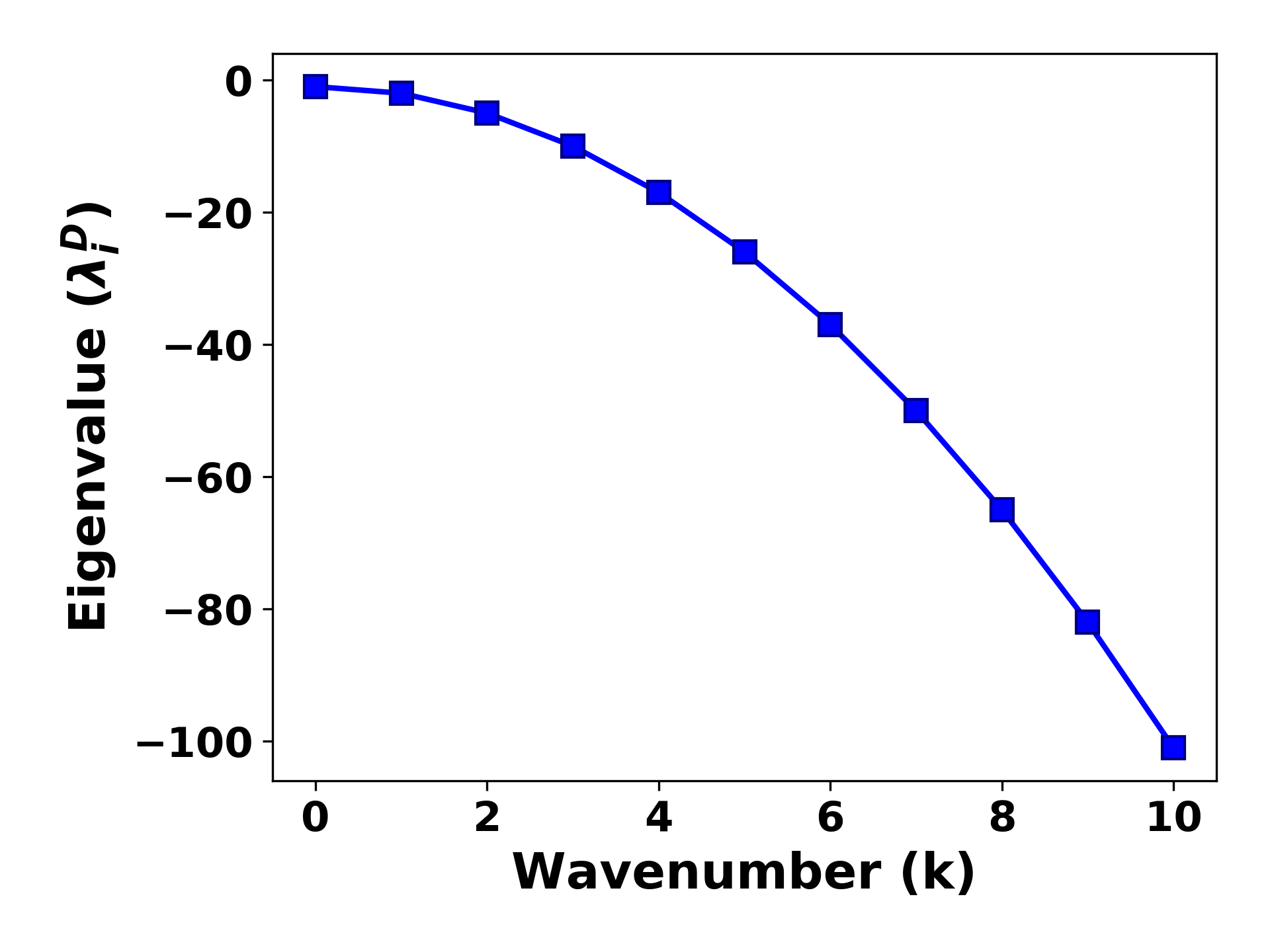}
    \caption{Eigenvalue $\lambda_i^D$ (given in Eq. \ref{lambda}) as a function of wavenumber $k$ of the Linear Stability Matrix $\mathcal{A}_{\mathcal{D}}$. We notice $\lambda_i^D$ to become more negative with the increase in $k$. This signifies that the system to become more and more stable with the increase in $k$. The parameters to obtain this figure are the same as Fig. 5a. For the other parameter sets, we obtain similar qualitative results, i.e. the eigenvalues becoming more and more negative with the wavenumber $k$.}
    \label{FFL_lambda}
\end{figure}
\section{Conclusion and Future directions}\label{diss}
Emergent responses in biological circuits as a consequence of context dependency have drawn the attention of the research community in the recent past \cite{tan2009emergent,nikolados2019growth,chakraborty2021emergent,9691638}. Among these several context dependencies, ribosome limitation is a major controlling factor in gene expression dynamics. Different processes, which are apparently not connected, get coupled implicitly due to the limited presence of this essential translational resource in the neighboring cellular environment. In this work, we have taken some commonly occurring motifs, specifically three gene patterns, which are quite different from conventional FFL structures as a repressive regulation is absent. The absence of this regulatory arm can arise from mutation in the system, which is a very random stochastic yet unavoidable change in genome structure. In humans, an average of 175 mutations per diploid genome per generation (i.e., the average mutation rate of $2.5\times 10^{-8}$) is noted \cite{nachman2000estimate}, while in \textit{E. coli} the average mutation rate is $2.1\times 10^{-7}$ per gene per generation  \cite{chen2013no}. The selected motifs show an FFL-like response when driven by resource competition, in a resource-limited cellular environment. Emergent repression arising from the context dependency, more precisely two mRNAs competing for the translational ribosome, fulfills the repression condition in the chosen motifs, and the fundamental functional responses of conventional FFLs, like response delay, pulse generation, dependency of steady sates upon inducers, etc., are achieved in proposed rFFL architectures. Acceleration or delay in response is depicted in the cFFL motif due to its unique construction, and the same is achieved in rFFL, solely caused by sharing resources from a common ribosome pool. This nonlinear coupling between the host and the circuit can modulate the dynamics of the entire system significantly. Our work not only depicts the possibilities of vast modification in gene circuit response due to resource limitation but also proposes an emergent architecture for one of the most common genetic motifs, feed-forward loops. It should also be reported that, in our rFFL motifs, we were able to achieve further complex patterns for different initial conditions. With the increasing complexity of the initial distribution of $g_x$ and $\beta_y$, we find complex patterns in $Z$ output.  Here. multiple stripe-forming patterns are connected together in order to get a complex pattern in rFFL motifs, which are stable through the simulation time for a wide range of diffusion coefficients.  
\\The patterns, we received from rFFL networks, not only signifies to the strength of our rFFL motifs in mimicking the cFFL topologies, establishing the strength of resource competition emerging new responses in the system, but also open up a new field of synthetic pattern formation. The synthetic biology approach to pattern formation has had major success in the recent past. Extreme parameter dependency of the Turing pattern, however, plays a major drawback in achieving the Turing pattern in synthetic biological systems solely \cite{karig2018stochastic,sekine2018synthetic}. So, a recent approach is gaining attention from synthetic biologists to achieve pattern formation in more than two node systems other than by classical parameter-sensitive Turing methodologies \cite{zheng2016identifying,diego2018key}. Some other recent models receive persistent transient spatiotemporal patterns in the system via diffusible protein molecules considering positional information in the modeling \cite{roy2023spatiotemporal,chakraborty2023quantitative,chakraborty2023spatio}. Our model supports a new mechanism of synthetic pattern formation, by controlling resource allocation in the nodes. Further, by setting the initial condition as a complex function,  complex patterns are achieved computationally. Further experimental verification will open up a new field of complex synthetic pattern formation. In our work, we have considered isotropic diffusion in a no-flux boundary condition. However, protein diffusion is not isotropic all the time but is dependent on various biological factors also. It is observed that diffusion can be different in different directions as well can be a gradient function too. This serves a large scope of exploring our proposed synthetic pattern formations for robustness and as well for novel pattern formations in the future.
\\It is important to note that our considerations are only valid for a resource-limited, low-growth system. The growth of the system is directly linked with the number of active ribosomes participating in translation and thus with the biomass of the system. Cellular
macromolecular composition could be highly correlated with cell growth \cite{klumpp2009growth}, and further investigations can be planned considering this factor in future work. Moreover, biological processes mostly take place in a noisy environment. In a  recent work on the noise characteristics of FFL \cite{ghosh2005noise}, the relation between functionality and abundance has been suggested, keeping the noise factor in mind. Deterministic and stochastic characteristics of functionality, dynamics, and response of the proposed rFFL motifs can also be elaborately studied, to develop a further understanding of complex synthetic circuit operation in diverse host cells.
\section*{Conflict of Interest}
The authors declare that they do not have any known conflicts of interest.

\section*{Data Availability}
\noindent The manuscript has no associated data.

\section*{Acknowledgement}
\noindent PC and SG acknowledge the support  by DST-INSPIRE, India, vide sanction Letter No. DST/INSPIRE/04/2017/002765  dated- 13.03.2019. UR acknowledges the support  by DST-INSPIRE, India, vide sanction Letter No. DST/INSPIRE/04/2022/003052  dated 27.02.2023.

\appendix

    \label{complex}
    \label{difcoef}
\newpage
\bibliography{aapmsamp}

\end{document}